\documentclass[amsmath,amssymb,twocolumn,aps,prb,superscriptaddress]{revtex4-1}

\usepackage{mathtools}
\usepackage{mathrsfs, amsmath, wasysym}
\usepackage[mathscr]{eucal}
\usepackage{xfrac}
\usepackage{graphicx}
\usepackage{dcolumn}
\usepackage{bm}
\usepackage{color}
\usepackage{tabularx}
\usepackage{hyperref}
\hypersetup{colorlinks,linkcolor=blue,citecolor=blue,urlcolor=blue}
\usepackage[normalem]{ulem}
\usepackage[all]{hypcap}
\usepackage[dvipsnames,usenames,table]{xcolor}
\usepackage{chngcntr}

\newcommand{\nocontentsline}[3]{}
\newcommand{\tocless}[2]{\bgroup\let\addcontentsline=\nocontentsline#1{#2}\egroup}

\newcommand{\bk}{{\bf k}}

\newcommand{\bq}{{\bf q}}

\newcommand{\be}{\begin{equation}}
\newcommand{\ee}{\end{equation}}
\newcommand{\beg}{\begin{gather}}
\newcommand{\eeg}{\end{gather}}
\newcommand{\beq}{\begin{eqnarray}}
\newcommand{\eeq}{\end{eqnarray}}
\newcommand{\bea}{\begin{align}}
\newcommand{\eea}{\end{align}}
\newcommand{\beqq}{\begin{eqnarray*}}
\newcommand{\eeqq}{\end{eqnarray*}}

\newcommand{\ket}[1]{ | #1 \rangle }

\begin{document}

\title{Controllable Topological Insulator Phases in Litharge-phase InBi Monolayer}

\author{Zhenyao Fang}
\affiliation{Department of Chemistry, University of Pennsylvania, Philadelphia, Pennsylvania 19104--6323, USA}

\author{Andrew M. Rappe}
\affiliation{Department of Chemistry, University of Pennsylvania, Philadelphia, Pennsylvania 19104--6323, USA}

\begin{abstract}
Despite recent advances of layered square-net topological material models that possess ideal semimetallic electronic structures and promising potential in material applications, the identification of experimentally accessible two-dimensional square-net materials with related topological properties has proven challenging. Due to the highly tunable physical and topological properties of III-V semiconductors, we revisit the class of III-V materials and observe that the litharge-phase InBi is a layered square-net material and can be exfoliated into the InBi monolayer. We present a comprehensive first-principles study of the energy landscape of the InBi monolayer. We identify a paraelastic phase and three ferroelastic phases and study their topological properties. Specifically, we show that the paraelastic InBi monolayer is a trivial insulator due to the orbital-ordering-induced band inversion occurring between states with the same parity. Substituting one Bi atom per cell with another V-group element (N, P, As) or applying an electric field that breaks the inversion symmetry and changes the orbital onsite energy, the paraelastic InBi monolayer can be driven into the topological insulator phase. Furthermore, one of the ferroelastic phases of pure InBi, which can be obtained by gently straining the paraelastic phase, also possesses such topological insulating properties. These results provide several experimentally accessible routes to tune the nontrivial topology in the InBi monolayer, including creating heterostructures with piezoelectric or ferroelectric substrates and applying mechanical strain, making the InBi monolayer an ideal platform to study the interplay of reduced dimensionality, square-net chemical bonding networks, and band topology.
\end{abstract}

\maketitle

\section{Introduction \label{sec:intro}}
The family of square-net topological materials, mostly topological semimetals \cite{Lee21p165106, Klemenz2019p185}, has been a center of attraction in the condensed matter community due to their ideal topological semimetallic electronic structures (predicted to give rise to optical conductivity linear to photon frequency as well as high carrier mobility) \cite{Klemenz20p13, Xian17p041003, Schilling17p187401, Topp17p041073}, coexistence between magnetic orders and charge density waves \cite{Lei19p1900045}, and exotic optical responses (such as second-order harmonic generation responses that are mainly contributed by surface states) \cite{Kirby22p838}. Various theoretical tight-binding models accurately predict the nodal line or nodal point electronic structures; the link between the topology and chemical bonding in this family of materials is well understood, providing multiple design strategies to finely control and study the profound Dirac and Weyl physics, such as by tuning the atomic distances, the spin-orbit coupling strength, or the electron filling \cite{Tremel87p124, Shan17p054003, Takane16p121108}.

However, most of the square-net materials discovered until now are three-dimensional (layered) materials \cite{Klemenz2019p185, Lee21p165106}. Therefore, the question naturally arises whether other nontrivial topological phases could arise when going to the two-dimensional limit, since two-dimensional materials often possess unique electronic properties that are absent in their three-dimensional counterparts \cite{CastroNeto09p109, Splendiani10p1271, Mak10p136805, Radisavljevic11p147}. Moreover, these properties are highly tunable by size and shape control \cite{Li19p1902431, Liscio17p025017}, layer-number control \cite{Wei22p1, Novoselov16p461, Cain20p26135}, strain effects \cite{Peng20p190, Sun19p082402, Bissett14p11124}, and gating modulation \cite{Cai18p21}, making these materials ideal platforms to study the effect of nontrivial topology on material properties. 



In order to integrate the unique properties of ultrathin materials into the family of square-net materials, it is important to find experimentally accessible topological material candidates. One class of materials which possesses such possibilities is the conventional III-V semiconductors, such as GaAs and InAs \cite{Johnson90p3655, Barrigon19p9170, Li18p1800005}. Apart from their profound and intriguing physical properties, they can also be tuned into various nontrivial topological phases by ion substitution or by strain modulation, making them ideal platforms to study the interplay between topology and solid-state materials. In the case of the zincblende phase \cite{Huang14p195105}, Bi substitution into GaAs, GaSb, and InSb can induce a band inversion and generate a topological semimetal similar to HgTe, which can then be transformed into a topological insulator phase after applying uniaxial strain. In the case of the wurtzite phase \cite{Fang20p125202}, varying the Bi concentration can tune GaAs into "near Dirac" triple-point semimetals (with only a pair of triple-points on the Fermi level) and Dirac-Weyl semimetals (with coexisting Dirac points and Weyl points \cite{Gao18p106404}). However, the strong covalent bonds in the conventional III-V crystal structures (zincblende and wurtzite structures) obstruct the experimental synthesis of ultra-thin layers by exfoliation methods, thus hindering their applications in the two-dimensional limit. Therefore, it is promising to identify two-dimensional square-net III-V material candidates with highly tunable topological properties.

In this work, we revisit the class of III-V materials and observe that the litharge phase InBi is a layered material with weak van der Waals interactions between the two Bi atomic planes in adjacent unit cells, contrary to other III-V materials. According to first-principles calculations, as the number of layers decreases, a ferroelastic (FE) litharge structure (which we call FE-I in the following) emerges as the dynamically stable structure, whereas the original paraelastic (PE) structure that derives from the bulk litharge phase becomes dynamically unstable, consistent with the previous study \cite{Ding2022p21546}; the energy surface becomes a double well, with the PE structure being the saddle point connecting the two FE-I phases. Apart from these two structures, we also reveal two other strongly buckled FE phases by calculating the energy landscape of InBi monolayer, which we call FE-II and FE-III. We also show that these PE and FE InBi monolayers (a square-lattice In atomic plane sandwiched by two Bi atomic planes) are either topologically trivial insulators or metals. Specifically, in the PE InBi monolayer, the band inversion occurs between the $p_{x,y}$ states and the $p_z$ states of Bi atoms with the same parity, which does not lead to a topologically nontrivial phase. Furthermore, as a result of elemental substitution such as As or P, the onsite orbital energy of one of the anion states decreases and the inversion symmetry is broken, splitting the doubly-degenerate bands and leading to a band inversion; this realizes PE InAs$_{0.5}$Bi$_{0.5}$, InP$_{0.5}$Bi$_{0.5}$, and InN$_{0.5}$Bi$_{0.5}$ as topological insulators. The mechanism leading to nontrivial topology is different from other two-dimensional topological insulators. In single-element monolayers (silicene, germanene, antimonene, bismuthene) \cite{Xian17p041003, Liu11p195430, Zhang15p3112}, the band inversion is driven by the spin-orbit coupling effect in the $p_z$ orbitals, and in functionalized monolayers BiX/SbX (X = H, F, Cl, Br) \cite{Liu14p085431} and buckled honeycomb III-V monolayers (such as GaBi, InBi, TlBi) \cite{Crisostomo15p6568}, it is driven by the spin-orbit coupling effect in the $p_{x,y}$ orbitals. In InAs$_{0.5}$Bi$_{0.5}$, InP$_{0.5}$Bi$_{0.5}$, and InN$_{0.5}$Bi$_{0.5}$, band inversion results from orbital energy ordering between different $p$ orbitals and from inversion symmetry breaking.Since topological properties are robust against weak perturbations, the FE-I alloys that can be obtained by applying weak strain to the PE phase are still topological insulators with a finite gap, but also electron and hole pockets, across the Brillouin zone.


\section{Methodology \label{sec:methods}}
Our calculations were performed based on density functional theory, as implemented in Quantum Espresso \cite{Giannozzi09p395502}. We used norm-conserving pseudopotentials generated by the OPIUM code\cite{Opium}, where for N atoms only valence electrons were treated explicitly, for P atoms empty 3d states were also included in the pseudopotential reference configuration as they receive a nonnegligible effect from empty d orbitals, while for other elements (In, As, Sb, Bi), filled semicore d-electrons were also included. We used GGA-PBE functionals \cite{Perdew96p3865} for all calculations, including structural relaxation and electronic property calculations; using hybrid functionals (such as HSE06 \cite{Paier06p154709}) does not change our conclusions qualitatively. We used the DFT-D3 method with Becke-Johnson damping to describe the vdW interactions \cite{Grimme06p1787, Grimme10p154104}.

For structural relaxations and electronic structure calculations, we used $16 \times 16 \times 1$ $\bk$-point sampling grid and 50~Ry for the kinetic energy cutoff for the plane-wave basis sets. For geometry optimization, the force convergence criterion of 0.005~eV/\AA, and a vacuum of thickness of 20 \AA{} were used. Furthermore, for two-dimensional InBi, we calculated its elastic constants (with independent components $C_{11}$ and $C_{12}$) by stress-strain methods, and the Young's modulus is defined as $E = (C_{11}^2 - C_{12}^2) / C_{11}$ \cite{Wei09p205407}. We also performed phonon dispersion relation calculations to study the thermodynamic stability of the monolayer, and they were calculated with a $6 \times 6 \times 1$ $ \mathbf{q}$-point grid and a kinetic energy cutoff of 120~Ry.

To study the topological properties of the proposed materials, we construct the tight-binding Hamilonian using maximally-localized Wannier functions, as implemented in Wannier90 \cite{Marzari12p1419, Mostofi14p2309}, with the projections chosen to be the s-orbital states of the In atoms and the p-orbital states of the anions (N, P, As, Sb, Bi). Furthermore, we use Wanniertools to calculate the $\mathbb{Z}_2$ index, Wannier charge center flow, and the surface states based on the Hamiltonian obtained from Wannier90 \cite{Wu18p405}. 

\section{Crystal structure \label{sec:crystal}}
\begin{figure}[htb]
\centering
\includegraphics[width=\linewidth]{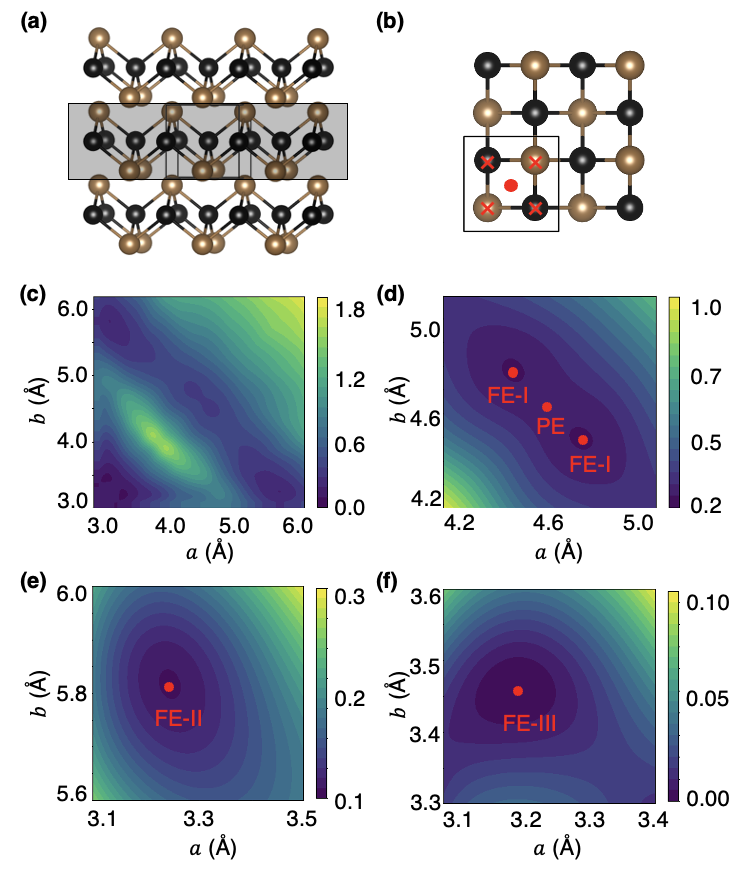}
\caption{(a) Side view of the crystal structure of litharge-phase InBi, where the black and brown atoms are In and Bi, respectively. The shaded area, containing a Bi-In$_2$-Bi trilayer, indicates the InBi monolayer. (b) Top view of the InBi monolayer, where the red dot and crosses represent the two choices of origin for space group symmetry operations. (c) The calculated energy landscape of the InBi monolayer, with energy (per unit cell) referenced to the FE-III structure, and the magnified view of the energy surface around the (d) PE and FE-I structures, (e) the FE-II structure, (f) and the FE-III structure.}.
\label{fig:structure_energy_landscape}
\end{figure}

Unlike III-V materials GaAs, GaSb, and InSb which energetically prefer the zincblende crystal structure, and GaN and InN which favor the wurtzite structure, InBi naturally crystallizes in the litharge (lead oxide) structure \cite{Gmitra16p165202, Ferhat06p115107}. Different from the zincblende and the wurtzite structure which feature strong covalent bonds between the anions and the cations, the litharge structure, shown in Fig.~\ref{fig:structure_energy_landscape}(a), can be viewed as Bi-In$_2$-Bi trilayers stacked in the $z$ direction \cite{Degtyareva98p9}. In this crystal structure, the In atoms are tetrahedrally coordinated with Bi atoms, each Bi atom is only tetrapodally coordinated (Bi with four adjacent In atoms forms a square pyramid). The discrepancy between the litharge structure and other conventional III-V crystal structures, where all ions are tetrahedrally coordinated with oppositely charged ions, arises due to the $6s^2$ electron lone pairs of Bi atoms which stabilize the litharge structure through a second-order Jahn-Teller distortion. Similar effects are also found in PbO, Bi$_2$Ti$_2$O$_7$, and CsSnBr$_3$ \cite{Walsh05p1422v2, Fabini16p11820v2, Melot09p224111}. In the litharge structure, adjacent Bi-In$_2$-Bi trilayers are coupled by weak van der Waals interactions; the Bi-Bi bonding strength is significantly smaller than those of In-In and In-Bi atomic pairs, as shown by the crystal orbital Hamilton population analysis (COHP) result in Appendix~\ref{app:COHP}. The exfoliation energy per surface area $A$ is $E_\text{exfoliation} = (E_\text{bulk} - E_\text{monolayer}) / (2A) = 27.4$ meV/\AA$^2$, similar to other well-known layered materials such as graphite (24.3 meV/\AA$^2$) \cite{Wang15p7853}, MoS$_2$ (26.3 meV/\AA$^2$) \cite{Fang20p23419}, and Bi$_2$Se$_3$ (22.4 meV/\AA$^2$) \cite{Melamed17p8472}. 

We performed calculations on the energy landscape of InBi monolayer and identified the high-symmetry PE phase and three other FE phases, indicated in Fig~\ref{fig:structure_energy_landscape}(c). Firstly, in the PE phase, within each Bi-In$_2$-Bi trilayer (which we call InBi monolayer in the following, shown as the shaded area in Fig.~\ref{fig:structure_energy_landscape}(a)), In atoms form a square lattice, sandwiched by Bi atoms from above and below. As a comparison, most previously reported InBi monolayers are formed by triangular lattices of In and Bi atoms located at two sublattices; these are higher in energy than litharge InBi monolayer by around 0.54~eV per formula unit and can only be dynamically stable by functionalization. \cite{Li16p23242, Barhoumi18p171, Zhang18p7022, Shen2022p207320}. After structural relaxation, the equilibrium lattice constant of the litharge InBi monolayer is $a(\text{PE})=4.62$~\AA, and the distance between Bi and In atomic planes is 1.94~\AA. The calculated Young's modulus of InBi monolayer is $E = 129.31$~GPa~\AA, much smaller than typical two-dimensional materials such as graphene, transition metal dichalcogenides, or h-BN \cite{Memarian15p348, Li18p49, Falin17p15815}. Secondly, we also identify a FE phase, which we call as FE-I phase in the following, near the PE phase, consistent with previous findings \cite{Ding2022p21546}. In this phase, the In atoms form a buckled rectangular lattice with the buckling distance (defined as the distance between adjacent In atoms in the $z$-direction) being 0.25~\AA. The spontaneous strain occurring in the FE-I phase makes it a lower-energy structure than the PE phase (see Tab.~\ref{tab:energy_landscape}). Furthermore, the phonon band structure of the FE-I InBi monolayer, shown in Appendix~\ref{app:phonon}, suggests that it is dynamically stable. This result, along with the exfoliation energy calculations indicating the weak van der Waals nature of the interlayer interactions, suggests the possibility of experimentally realizing this monolayer through exfoliation methods. Finally, we identified two other FE phases (which we call as FE-II and FE-III phases in the following) that are strongly compressed along one in-plane axis, and the In atoms are strongly buckled. Their lattice constants and buckling distances are summarized in Tab.~\ref{tab:energy_landscape}.

\begin{table}[htb]
    \centering
    \begin{tabular}{|c|c|c|c|c|}
    \hline
    & $a$~(\AA) & $b$~(\AA) & $d_\text{(buckle)}$~(\AA) & $E$ (eV/cell)\\ \hline
    PE & 4.62 & 4.62 & 0.00 & -13.97 \\ \hline
    FE-I & 4.79 & 4.45 & 0.25 & -14.00 \\ \hline
    FE-II & 5.80 & 3.26 & 1.68 & -14.17 \\ \hline
    FE-III & 3.19 & 3.46 & 2.32 & -14.30 \\ \hline
    \end{tabular}
    \caption{The lattice constants $a$ and $b$, the buckling distance (defined as the distance between the In atoms in the $z$-direction), and the energy of the PE, FE-I, FE-II, FE-III phases of the InBi monolayer.}
    \label{tab:energy_landscape}
\end{table}

The symmetry group of the PE InBi monolayer (space group $P4/nmm$ (No. 129) and the associated point group ($D_{4h}$)) are generated by the following operations: four-fold screw rotational symmetry $\bar{C}_{4z}=C_{4z} t(\frac{1}{2}00)$, two-fold screw rotational symmetry along the $z$-axis $\bar{C}_{2z}=C_{2z} t(\frac{1}{2}\frac{1}{2}0)$, two-fold screw rotational symmetry along the $y$-axis $\bar{C}_{2y}=C_{2y} t(0\frac{1}{2}0)$, and inversion symmetry. Note that the inversion center is placed at the midpoint of In-In bonds (indicated as the dots in Fig.~\ref{fig:structure_energy_landscape}(b)), instead of at the centers (right below or above a Bi atom) or the corners of In atomic squares (In atoms), indicated as the crosses in Fig.~\ref{fig:structure_energy_landscape}(b). With this choice of origin (dots), rotational symmetries become screw axes. On the other hand, if we choose any of the crosses as the origin, the rotational symmetries along the z-axis remain as pure rotational, while the inversion symmetry must be modified to $\bar{i} = i t(\frac{1}{2}\frac{1}{2}0)$. 

For the other three FE phases, the four-fold screw rotational symmetry $\bar{C}_{4z}$ is broken, reducing the space group symmetry to $Pmnm$ (No. 59) and the associated point group to $D_{2h}$. These three phases share the same symmetry group, and only differ in the lattice constants and the buckling distance.

\section{Results and Discussions \label{sec:results}}
\subsection{PE InBi Monolayer}

\begin{figure*}[htb]
\centering
\includegraphics[width=\linewidth]{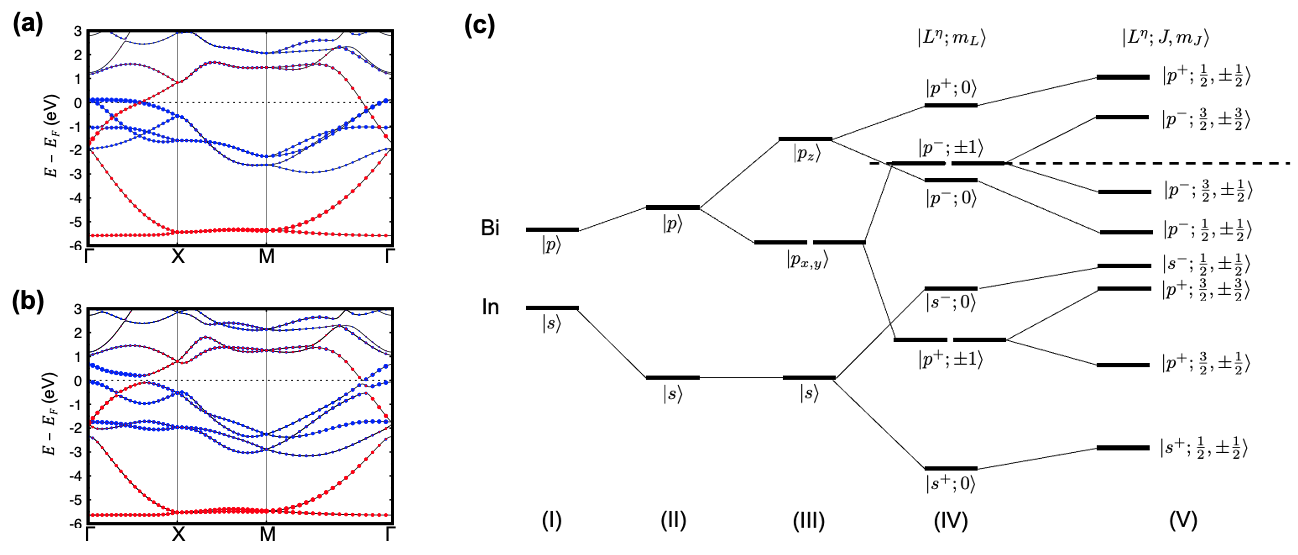}
\caption{The calculated electronic structures of PE-phase InBi monolayer without (a) and with (b) spin-orbit coupling, where red and blue circles represent the projections onto the s-orbitals of In atoms and the p-orbitals of Bi atoms, respectively. (c) The orbital evolution diagram of InBi monolayer under the effects of In-Bi interactions (II), crystal field splitting (III), next nearest neighbor hopping (the In(Bi) states forming electronic states with definite parity with four next nearest In(Bi) neighbors, indicated as IV), and spin-orbit coupling (V), where the black dotted line indicates the Fermi level.}.
\label{fig:InBi_bands}
\end{figure*}

The litharge bulk phase InBi is a Dirac nodal line semimetal without spin-orbit coupling. Once the spin-orbit coupling is included, the Dirac nodal lines split, leaving several Dirac points at high symmetry points or along high-symmetry lines in the Brillouin zone, as shown in Appendix~\ref{app:COHP}. With the presence of other trivial bands at similar energy to the Dirac points, it may be difficult to experimentally detect the properties associated with the Dirac points in conductivity or light absorption, making it a less than ideal Dirac semimetal. However, the metallic nature distinguishes litharge phase InBi from other conventional III-V semiconductors, which can be easily tuned into various topologically nontrivial phases by strain engineering or by ion substitution; this difference, as described above, is also reflected in their crystal structures.

For the PE InBi monolayer that directly derives from the bulk InBi by exfoliation, the calculated electronic structures are shown in Fig.~\ref{fig:InBi_bands}(a, b). Without spin-orbit coupling, it is also a Dirac nodal line semimetal. Once spin-orbit coupling is included, these nodal lines are gapped out, making it a semiconductor with a band gap of 0.1 eV; this is different from its three-dimensional counterpart which is a Dirac semimetal. Similar to other III-V materials, the bands around the Fermi level mainly comprise the s-orbitals of In and p-orbitals of Bi that are higher in energy, shown as step (I) in Fig.~\ref{fig:InBi_bands}(c). Without spin-orbit coupling, the orbital evolution (at $\Gamma$ point) under the crystal-field splitting effect is shown as step (II) in Fig.~\ref{fig:InBi_bands}(c). Then due to the presence of inversion symmetry, these orbitals can form bonding and antibonding states with definite parity $\eta=\pm 1$ \cite{Liu10p045122}, defined as $\ket{s^\pm} = \frac{1}{\sqrt{2}} (\ket{\text{In1}, s} \pm \ket{\text{In2}, s})$ and $\ket{p_\alpha^\pm} = \frac{1}{\sqrt{2}} (\ket{\text{Bi1}, p_\alpha} \mp \ket{\text{Bi2}, p_\alpha})$, where $\alpha=x,y,z$, the superscript $\pm$ indicates the parity, and In1(Bi1) and In2(Bi2) are related to each other by inversion symmetry. From Fig.~\ref{fig:structure_energy_landscape}(b), each In1(Bi1) atom is actually related to four other In2(Bi2) atoms by inversion symmetry (note that an inversion center is found at the center of each square), so here In2(Bi2) actually represents the linear combination of all these four In2(Bi2) states and we can view this step (III) as next nearest neighbor hopping. Furthermore, the $\ket{p_x^\pm}$ and $\ket{p_y^\pm}$ orbitals can form states with definite $z$-component of orbital angular momentum $m_L$, allowing us to label the states by both parity and $m_L$. Using the notation $\ket{p^\eta; m_L}$, we defined $\ket{p^\pm; \pm 1} = \ket{p_x^\pm} \pm i \ket{p_y^\pm}$. The $\ket{p_z^\pm}$ orbitals themselves have orbital angular momentum $m_L = 0$, and thus we denote them as $\ket{p^\pm; 0}$. The orbital evolution of this hopping process is shown as step (III) in Fig.~\ref{fig:InBi_bands}(c). It is clear that although band inversion is present, it occurs between states $\ket{p^-; 0}$ and $\ket{p^-; \pm 1}$ with the same parity, which does not lead to a topologically nontrivial phase. 

The spin-orbit coupling effect (step IV) will couple the spin and orbital angular momenta, producing states with definite z-component of total angular momentum. Using the notation $\ket{L^\eta; J, m_J}$, the states are labeled as $\ket{s^\pm; \frac{1}{2}, \pm\frac{1}{2}}$, $\ket{p^\pm; \frac{3}{2}, \pm\frac{3}{2}}$, $\ket{p^\pm; \frac{3}{2}, \pm\frac{1}{2}}$ and $\ket{p^\pm; \frac{1}{2}, \pm\frac{1}{2}}$. By comparing with the calculated electronic structures without spin-orbit coupling, it is clear that spin-orbit coupling does not alter the orbital ordering or cause the band inversion, but only opens the band gap. Therefore, although it turns InBi monolayer into a semiconductor, spin-orbit coupling does not make InBi a topological insulator. This is also confirmed by calculating the Wannier charge center flow and the $\mathbb{Z}_2$ index of the InBi monolayer using maximally localized Wannier functions.

In order to further understand the effect of crystal field splitting and nearest neighbor hopping on band inversion, we also compared the electronic structures of PE InBi monolayer with InAs and InSb, if they were to crystallize in the same monolayer structure. To this end, we defined $E_\text{crystal field} = (E_{\ket{p_z^+}} + E_{\ket{p_z^-}}) / 2 - (E_{\ket{p_{x,y}^+}} + E_{\ket{p_{x,y}^-}}) / 2$ as the crystal field splitting strength (step III), $E_\text{NNN} = (|E_{\ket{p^+; 0}} - E_{\ket{p^-; 0}}| + |E_{\ket{p^+; \pm 1}} - E_{\ket{p^-; \pm 1}}|) / 2$ as the next nearest neighbor hopping strength (step IV), and $E_\text{inv} = E_{\ket{p^-; \pm 1}} - E_{\ket{p^-; 0}}$ as the magnitude of band inversion. The calculated electronic structures of InAs and InSb are shown in Appendix~\ref{app:alloy}. Firstly, even with spin-orbit coupling, InAs and InSb monolayers are still metallic, as the spin-orbit coupling of As and Sb is not strong enough to open a gap. Secondly, the above quantities of these materials, which characterize the effect of crystal field splitting and nearest neighbor hopping, are listed in Table~\ref{tab:band-quantities}. Since As is a smaller ion than Bi, the distance between the In atomic planes and the anion atomic planes becomes smaller (InBi: 1.94 \AA, InSb: 1.83 \AA, InAs: 1.69 \AA), leading to larger crystal field splitting. On the other hand, the nearest neighbor hopping strength differs little among these three materials. Based on the orbital evolution diagram, a larger crystal field splitting, or a smaller nearest neighbor hopping can lead to smaller band inversion. Therefore, by combining these two effects, among which crystal field splitting plays a major role, the magnitude of band inversion of InBi is the largest one (note that this does not come from the largest spin-orbit coupling effects of Bi atoms). 

\begin{table}[htb]
    \centering
    \begin{tabular}{|c|c|c|c|}
    \hline
    & $E_\text{crystal field}$ & $E_\text{NNN}$ & $E_\text{inv}$ \\ \hline
    InAs & 1.70 & 2.18 & 0.53 \\ \hline
    InSb & 1.33 & 2.16 & 0.95 \\ \hline
    InBi & 0.97 & 2.08 & 1.16 \\ \hline
    \end{tabular}
    \caption{The crystal field splitting strength $E_\text{crystal field}$, the next nearest neighbor hopping strength $E_\text{NNN}$, and the magnitude of band inversion $E_\text{inv}$ of PE InAs, InSb, and InBi monolayers}
    \label{tab:band-quantities}
\end{table}

\subsection{PE InAs$_{0.5}$Bi$_{0.5}$ Monolayer} \label{sec:alloy_elec}
To induce band inversion that leads to nontrivial topology, we revisited the ion substitution strategy, and replaced one of the Bi planes with elements in the same main group, shown in Fig.~\ref{fig:alloy_topo}(a). By comparing the electronic structures of InBi with InAs and InSb (shown in Appendix~\ref{app:alloy}), it is clear that the energy difference between the p-states of the anions and the s-states of the In atoms decreases from Bi to As. Therefore, ion substitution causes one set of the anion p-states to move to lower energy and also breaks the inversion symmetry. In the absence of inversion symmetry, the electronic states are no longer required to be doubly degenerate along the high-symmetry lines, opening the possibility of band inversion along these high-symmetry lines. 

Based on these considerations, we calculated the electronic band structures of InN$_{0.5}$Bi$_{0.5}$, InP$_{0.5}$Bi$_{0.5}$, InAs$_{0.5}$Bi$_{0.5}$, and InSb$_{0.5}$Bi$_{0.5}$ (shown in Appendix~\ref{app:alloy}). Among them, InN$_{0.5}$Bi$_{0.5}$, InP$_{0.5}$Bi$_{0.5}$ and InAs$_{0.5}$Bi$_{0.5}$ are topological insulators, while InSb$_{0.5}$Bi$_{0.5}$ is a trivial insulator. Therefore, in the following, we will mainly consider the electronic structure of InAs$_{0.5}$Bi$_{0.5}$ as an example, and it is shown in Fig.~\ref{fig:alloy_topo}(b). Since the inversion symmetry is broken, we cannot obtain the $\mathbb{Z}_2$ invariant from the parity eigenvalues, but only from calculating the flow of Wannier charge centers. As shown in Fig.~\ref{fig:alloy_topo}(c), we confirmed that the InAs$_{0.5}$Bi$_{0.5}$ monolayer is a topological insulator, and the associated surface states are shown in Fig.~\ref{fig:alloy_topo}(d). 

\begin{figure}[htb]
\centering
\includegraphics[width=\linewidth]{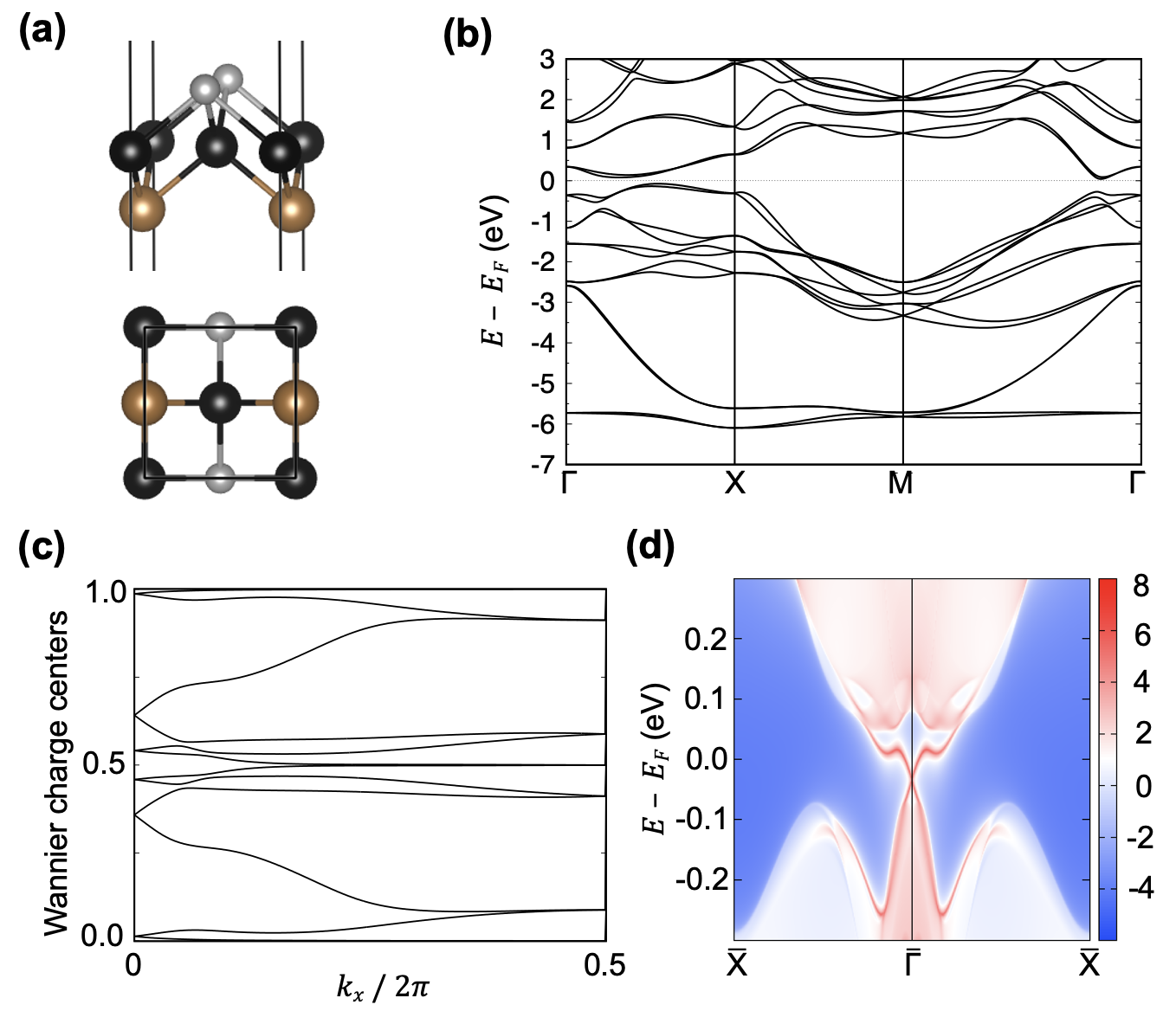}
\caption{(a) The side and top view of the alloy InAs$_{0.5}$Bi$_{0.5}$, where the black, brown, and gray atoms are In, Bi, As, respectively. (b) The calculated electronic structure of InAs$_{0.5}$Bi$_{0.5}$ with spin-orbit coupling. (c) The flow of the hybrid Wannier charge centers of InAs$_{0.5}$Bi$_{0.5}$. (d) The calculated (001) surface states of InAs$_{0.5}$Bi$_{0.5}$.}.
\label{fig:alloy_topo}
\end{figure}

\begin{figure*}[htb]
\centering
\includegraphics[width=\linewidth]{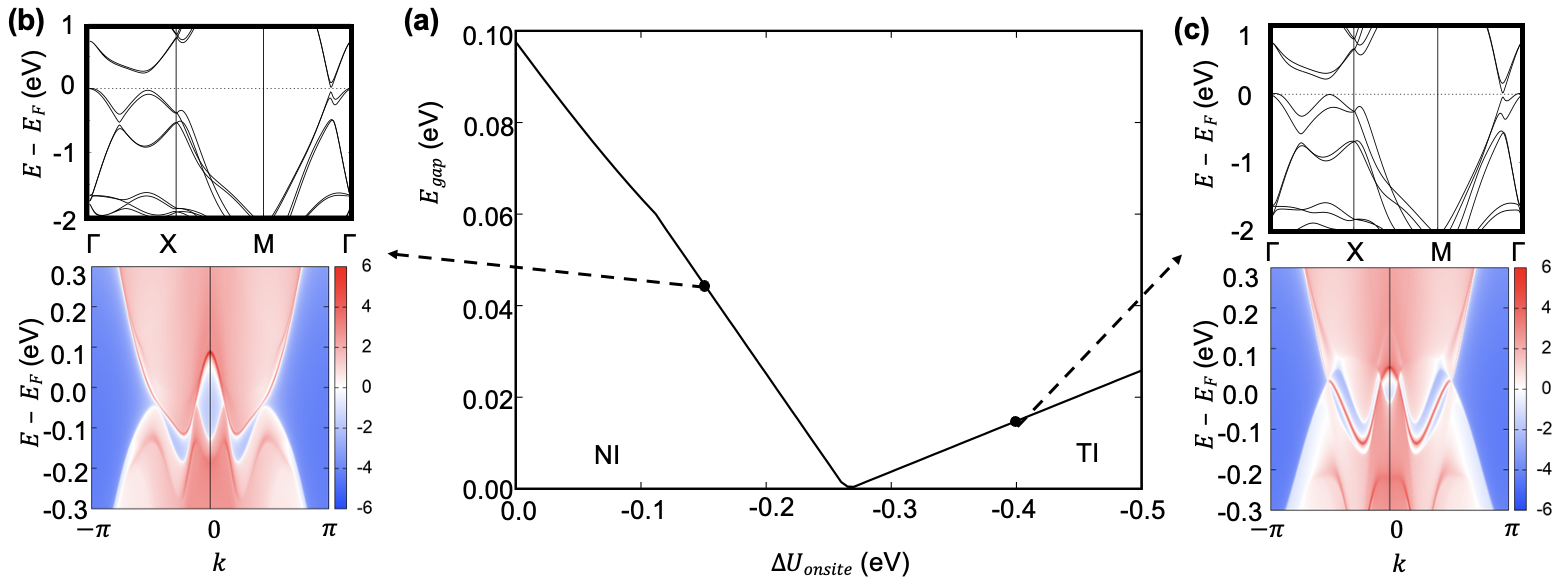}
\caption{(a) The calculated direct band gap of the InBi monolayer as the onsite potential of one Bi plane gradually decreases. The band structures and surface states with $\Delta U_\text{onsite} = -0.15$ eV and $-0.4$ eV are shown in panels (b) and (c) as examples.}.
\label{fig:gap}
\end{figure*}

As discussed above, spin-orbit coupling effect does not play a major role in driving the topological phase transition. Therefore, when replacing half the Bi with other V-group elements, the onsite potential of the p-states of those anions changes, as do the hopping parameters between these p-states and the s-states of the In atoms (due to smaller atomic distance). In order to investigate the effect of inversion symmetry breaking and further understand why InAs$_{0.5}$Bi$_{0.5}$ becomes a topological insulator, we return to the InBi case and tune the onsite energy of the anion states in the Wannier tight-binding Hamiltonian. By decreasing the energies of the Wannier functions corresponding to the p-orbitals of one Bi atom in InBi monolayer gradually, we observe successive band inversions. The first one, as shown in Fig.~\ref{fig:gap}, happens along the $\Gamma-M$ high-symmetry line, when the onsite energy decrease reaches 0.26~eV. This inversion causes the system to undergo phase transition from trivial insulator to topological insulator; band structures and associated surface states with energy decrease of 0.15 eV and 0.4 eV are shown in Fig.~\ref{fig:gap} as examples. When $\Delta U_\text{onsite} = -0.15$ eV, the Hamiltonian is semimetallic with electron pockets at $k = (0.25, 0.06, 0.00) \pi$ and hole pockes at $k = (0.11, 0.11, 0.00) \pi$; however, the direct band gap is still finite. By calculating its $\mathrm{Z}_2$ index and surface states, shown in Fig.~\ref{fig:gap}(b), it is topologically trivial. Besides, when $\Delta U_\text{onsite} = -0.40$ eV, we find it to be a topological insulator with nontrivial surface states, as shown in Fig.~\ref{fig:gap}(c). By calculating the band splitting between the top valence bands along the $\Gamma-X$ high symmetry line, we find that the splitting in the model Hamiltonian with $\Delta U_\text{onsite} = -0.40$ eV (0.34 eV) is similar to that of InAs$_{0.5}$Bi$_{0.5}$ (0.24 eV)., These results suggest that as the p-orbital onsite energy difference decreases (i.e., replacing one Bi atom with Sb, As, P, and then N atom), the monolayer will gradually change from a trivial insulator (which is the case of InSb$_{0.5}$Bi$_{0.5}$) to a topological insulator (which are the cases of InAs$_{0.5}$Bi$_{0.5}$, InP$_{0.5}$Bi$_{0.5}$, and InN$_{0.5}$Bi$_{0.5}$).

Apart from the onsite potential energy difference, other effects that break the inversion symmetry can also lead to the topological phase transitions, such as the hopping strength between the anions and cations (reflecting the bond length changes during the alloy process), the crystal field splitting within the $p$-orbitals of the anions, and the spin-orbit coupling strength. However, effects which preserve the $z$-direction isotropy, such as the biaxial compression, does not lead to topological insulating states (see Appendix~\ref{app:alloy}).

The above analysis not only explains the band inversion mechanism in InAs$_{0.5}$Bi$_{0.5}$, InP$_{0.5}$Bi$_{0.5}$, and InN$_{0.5}$Bi$_{0.5}$, but also provides additional experimentally accessible routes to realize nontrivial topological physics in the InBi monolayer. For example, the onsite potential can also be changed by applying an electric field in the direction normal to the monolayer. We calculated, through first-principles methods, that applying the electric field can even drive InBi into the topological insulator phase; the critical electric field for the phase transition is around 0.25 V/\AA. Another possible experimental method is to a heterostructure with a responsive substrate, such as piezoelectric or electrostrictive materials, such that by straining the substrate the amount of band inversion could be finely tuned. 

\subsection{FE-Phase InBi and InAs$_{0.5}$Bi$_{0.5}$ Monolayers} \label{sec:FE}
\begin{figure}[htb]
\centering
\includegraphics[width=\linewidth]{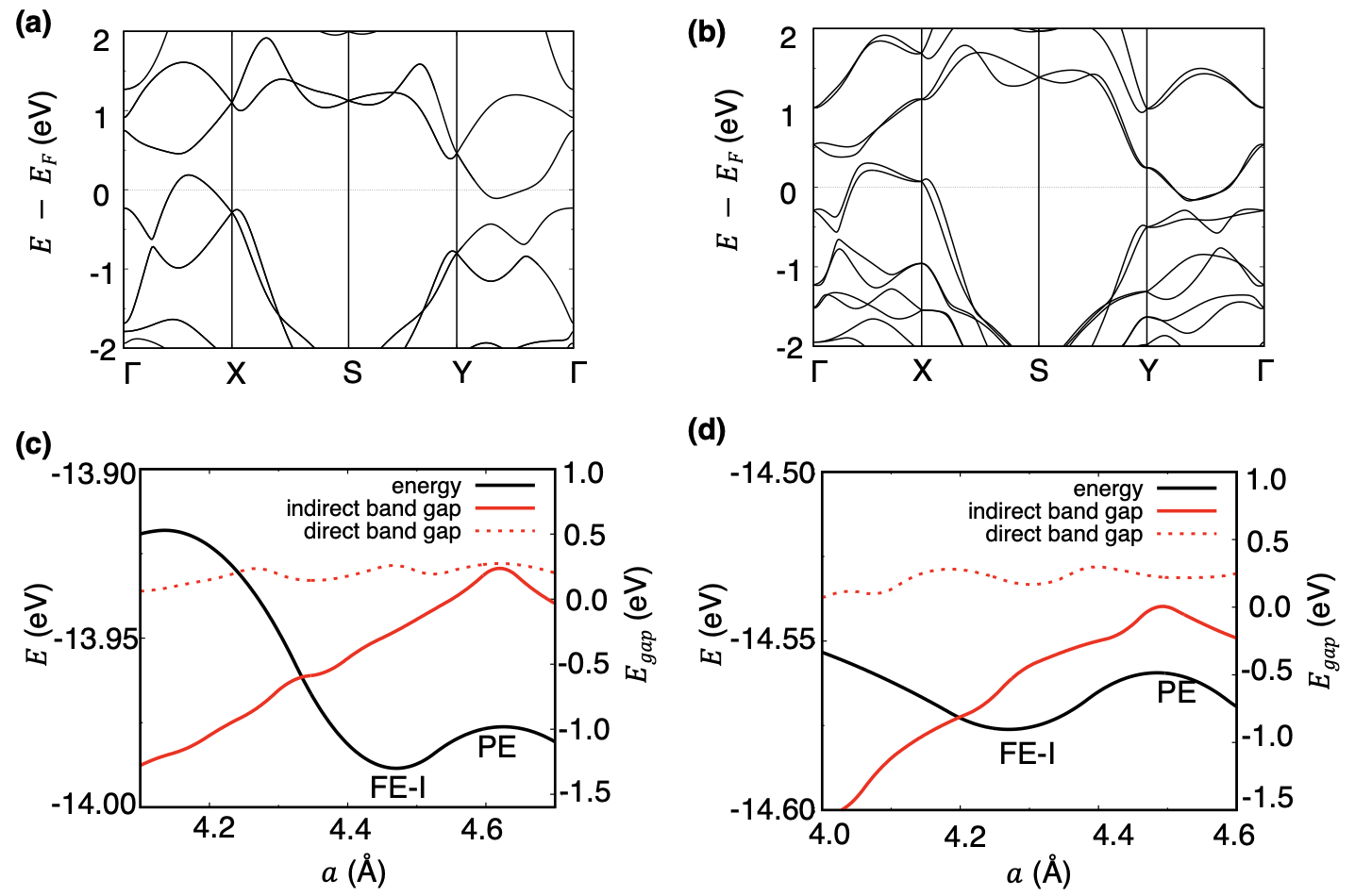}
\caption{(a, b) The calculated band structures of (a) FE-I InBi and (b) InAs$_{0.5}$Bi$_{0.5}$ monolayer along the high-symmetry lines. (c, d) The evolution of the total energy and the (indirect and direct) band gaps of InBi (c) and InAs$_{0.5}$Bi$_{0.5}$ monolayer (d) as the cell is compressed uniaxially. The solid black line is the total energy per unit cell, the solid red line is the indirect band gap, and the dashed red line is the direct band gap.}.
\label{fig:FE}
\end{figure}

We further analysed the effect of structural distortions (compression along one in-plane axis and buckling between adjacent In atoms) in the FE phases. Among these phases, the FE-I phase is the nearest to the PE phase. Because topological properties are robust against weak perturbations that do not close the band gap, we expect that the topological properties of FE-I InBi and alloy monolayers remain unchanged from their PE structures. Therefore, we first calculate the band structures of FE-I InBi and InAs$_{0.5}$Bi$_{0.5}$ monolayers, shown in panels (a) and (b) in Fig.~\ref{fig:FE}. Both monolayers have a finite band gap across the Brillouin zone, but the indirect band gaps become negative, indicating that electron and hole pockets emerge at different regions in the Brillouin zone.

Furthermore, we calculated the evolution of the (indirect and direct) band gaps of both monolayers as the lattice vector $a$ becomes shorter (while relaxing the lattice vector $b$), and plot it in panels (b) and (d) Fig~\ref{fig:FE}. In both monolayers, the PE phase has the largest positive band gap, and when going towards the FE-I phase, the indirect band gap turns negative while the direct band gap remains positive, suggesting no topological phase transitions occurring when compressing the PE phase to the FE-I phase. By calculating the $\mathbb{Z}_2$ indices, we verified that FE-I InBi is topologically trivial while FE-I InAs$_{0.5}$Bi$_{0.5}$ is nontrivial, consistent with the above discussions on the band gap evolution.

Finally, we also calculated the band structures of the FE-II and the FE-III InBi and InAs$_{0.5}$Bi$_{0.5}$ monolayers, shown in Appendix~\ref{app:FE}. These structures are all metallic and topologically trivial. These results demonstrate the possibility of finely tuning the topological properties in InBi and alloy monolayers by mechanical control.

\section{Conclusions}

In this work, we examined the structural phase diagrams and the electronic properties of the experimentally-accessible litharge-phase InBi monolayer and proposed different routes to tune the topological properties within this material. Based on first-principles calculations, we identified the PE phase and three FE phases of the InBi monolayer. By orbital analysis, the band inversions in PE monolayer InBi occur between electronic states with the same parity, thus making PE InBi monolayer a trivial insulator. However, by applying electric field in the vertical direction, or by alloying with other main group-V elements (N, P, As), PE InBi monolayer will become a topological insulator, because of the anion orbital onsite energy difference and the broken inversion symmetry that split the doubly degenerate bands and lead to band inversion along the $\Gamma-M$ high-symmetry line. 

Furthermore, the phase transition from the PE phase to the FE-I phase, although creating electron and hole pockets in different regions across the Brillouin zone, does not close the band gap and thus affect the topological properties. Therefore, the FE-I InBi monolayer is still topologically trivial while the alloys are nontrivial. Finally, the other two FE phases (FE-II, FE-III) are metallic and thus topologically trivial. The litharge-phase InBi and alloy monolayers exhibit profound structural phases and highly tunable topological properties, making them excellent experimental platforms for studying the diverse properties that arise from topology through the application of external fields or mechanical strain.

\section{Acknowledgements}
We acknowledge Tan Zhang for helpful discussions and Danna Freedman for sharing interest in bulk litharge InBi. This work was supported by the U.S. Department of Energy, Office of Science, Basic Energy Sciences, under Award No. DE-FG02-07ER46431. Computational support was provided by the National Energy Research Scientific Computing Center (NERSC).

\appendix
\counterwithin{figure}{section}
\counterwithin{table}{section}
\section{Crystal Orbital Hamilton Population Analysis of Litharge-Phase InBi} \label{app:COHP}
To understand the bonding character of litharge-phase bulk InBi, we performed the crystal orbital Hamilton population (COHP) analysis using the program LOBSTER \cite{Dronskowski93p8617}. By partitioning the density of states into chemical-bond-weighted density of states, this method allows us to understand the contributions of each type of chemical bonds to the electronic states near the Fermi level, and thus the chemical nature of the material. 

\begin{figure}[htb]
\centering
\includegraphics[width=\linewidth]{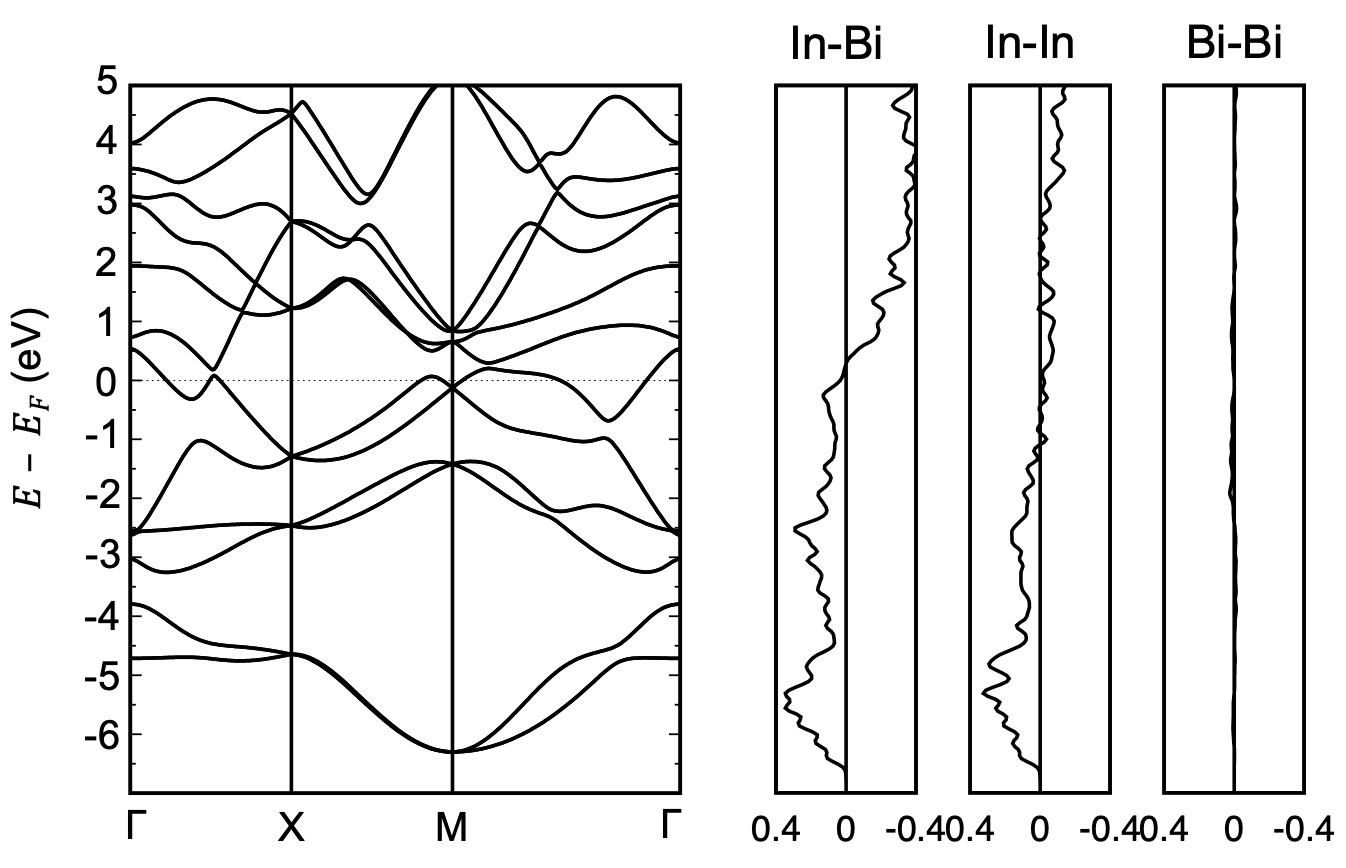}
\caption{The calculated band structure of three-dimensional litharge-phase InBi, and the crystal orbital Hamilton population for In-Bi, In-In, and Bi-Bi atomic pairs, where Bi-Bi pairs are the interlayer atomic pairs.}.
\label{fig:COHP}
\end{figure}

The calculated electronic structure (with spin-orbit coupling) and the COHP of litharge-phase bulk InBi are shown in Fig~\ref{fig:COHP}, where we focused on the chemical bonds In-Bi, In-In, and Bi-Bi. Positive and negative values COHP indicate the bonding and antibonding character for each type of bonds, respectively. From Fig~\ref{fig:COHP}, it is clear that the Bi-Bi bond strength is much smaller than those of In-Bi bonds and In-In bonds. Along with the fact that the calculated exfoliation energy of InBi is similar to those of other well-known layered materials, this result suggests that the interactions between Bi atoms belonging to vertically adjacent unit cells are very weak and mostly likely vdW interactions. This also indicates the necessity of using vdW correction schemes in first-principles calculations in order to predict the crystals structure of litharge-phase InBi accurately.

\section{Stability of the InBi Monolayer} \label{app:phonon}
\begin{figure}[htb]
\centering
\includegraphics[width=\linewidth]{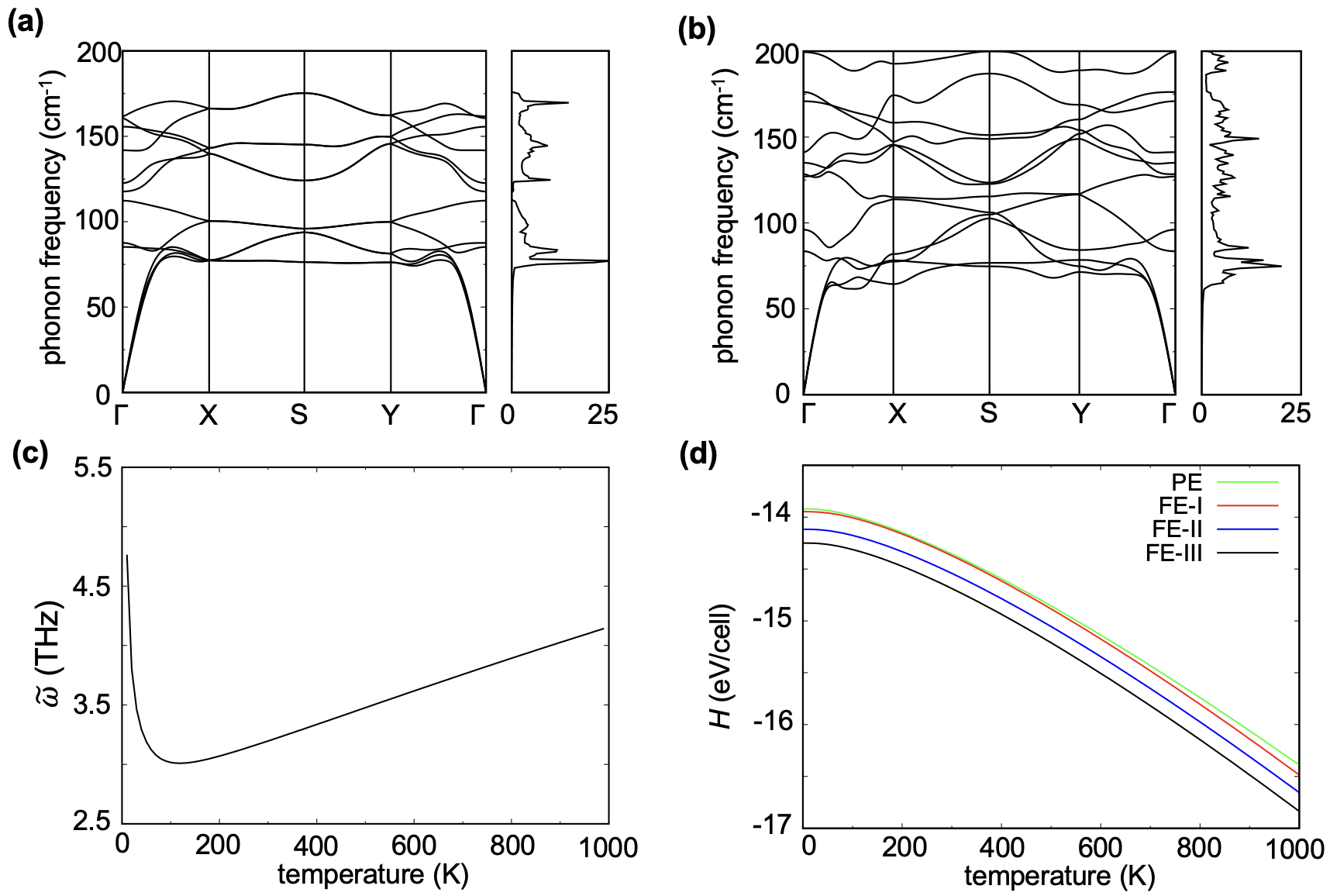}
\caption{(a) The calculated phonon dispersion relations and the density of states of FE-I InBi and (b) those of FE-I InAs$_{0.5}$Bi$_{0.5}$ monolayers. (c) The effective frequency of the imaginary phonon mode in the PE InBi. (d) The calculated Helmholtz free energy of PE InBi.}.
\label{fig:phonon}
\end{figure}

In order to study the dynamical stability of the InBi litharge monolayer, we calculated the phonon dispersion relations, shown in Fig.~\ref{fig:phonon}(a, b), on an $6 \times 6 \times 1$ $\bq$-grid. After enforcing the acoustic sum rules, the phonon dispersion relation is everywhere real, showing that the both FE-I monolayers are dynamically stable. Specifically, the fact that FE-I InBi is stable is consistent with the phonon calculations and the \emph{ab initio} molecular dynamics simulations (at 300 K up to 9 ps) in previous literature \cite{Ding2022p21546}.

Furthermore, we calculate the Helmholtz free energy to study the stability of the four phases of InBi monolayer. The phonon dispersions of the three FE phases don’t contain any imaginary phonon modes, indicating that they are all dynamically stable and that their vibrational entropy is well-defined. However, since the PE phase contains imaginary modes \cite{Ding2022p21546}, we calculate its free energy by renormalizing the imaginary modes \cite{Skelton16p075502}. We fit the double well potential (formed by the PE and the FE-I phase) by a polynomial, and calculate the partition function from the eigenvalues of the Hamiltonian. From the partition function we can derive the effective harmonic frequency through

\begin{equation}
\Tilde{\omega} (T) = \frac{2 k_B T}{\hbar}  \sinh^{-1} (\frac{1}{2Z(T)})   
\end{equation}

This effective frequency of the imaginary mode, shown in Fig.~\ref{fig:phonon}(c) at a target temperature T will reproduce its contribution to the thermodynamic partition function and to the Helmholtz free energy. From the Helmholtz free energy calculations shown in Fig.~\ref{fig:phonon}(d), the FE-III will remain as the ground state at higher temperatures.

Although the FE-I phase is not the ground state based on our calculations on the energy landscape of InBi monolayer (Fig.~\ref{fig:structure_energy_landscape}) and the Helmholtz free energies of the FE phases (Fig.~\ref{fig:phonon}), we find that there is an energy barrier between the FE-I and FE-II phases (0.08 eV/cell) and between the FE-I and FE-III phase (1.42 eV/cell). Therefore, at higher temperature it’s unlikely that the FE-I phase will go through the phase transitions towards the FE-II and FE-III phases. This is also confirmed in previous \emph{ab initio} molecular dynamics simulations \cite{Ding2022p21546}. On the other hand, since the energy difference between the FE-I and the PE phases is 0.03 eV/cell (they form a double well potential), phase transitions could occur between them. However, since both FE-I and PE InAs$_{0.5}$Bi$_{0.5}$ are topological insulators, the topological insulating properties will remain at higher temperature.
 
\section{Electronic Structures of PE InAs, InSb, and Alloy Monolayers} \label{app:alloy}
\begin{figure}[htb]
\centering
\includegraphics[width=\linewidth]{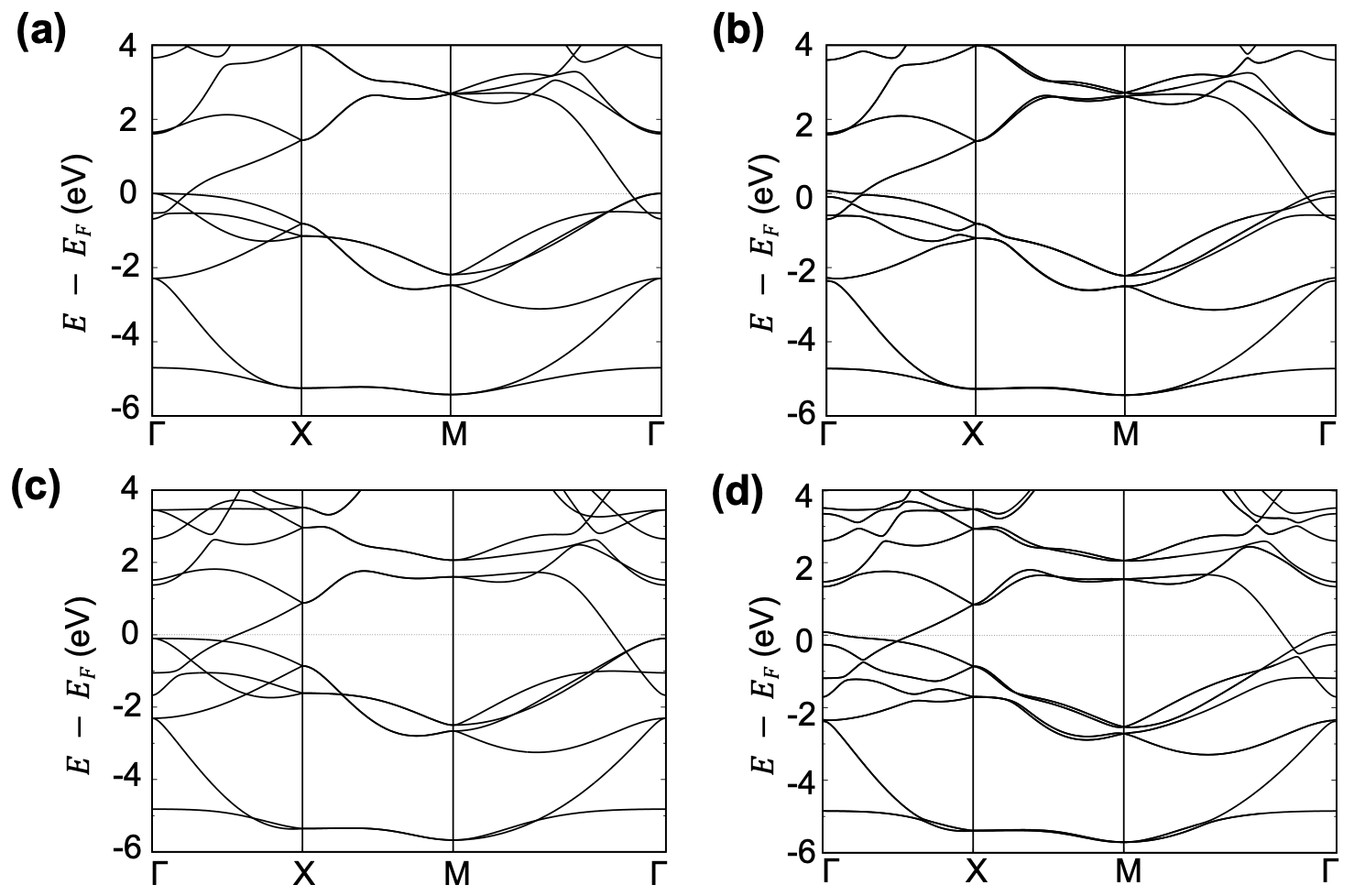}
\caption{The calculated band structures of (a) PE InAs without spin-orbit coupling and (b) with spin-orbit coupling, and (c, d) those of PE InSb (c, d).}.
\label{fig:InAsSb}
\end{figure}

As a comparison to the InBi monolayer, we also calculate the electronic structures of PE InAs and InSb monolayers to analyze the effect of crystal-field splitting and next-nearest-neighbor hopping. The calculated lattice constant of InAs is $a = 4.41$ \AA, and that of InSb is $a = 4.58$ \AA. Their calculated band structures are shown in Fig.~\ref{fig:InAsSb}.

\begin{figure}[htb]
\centering
\includegraphics[width=\linewidth]{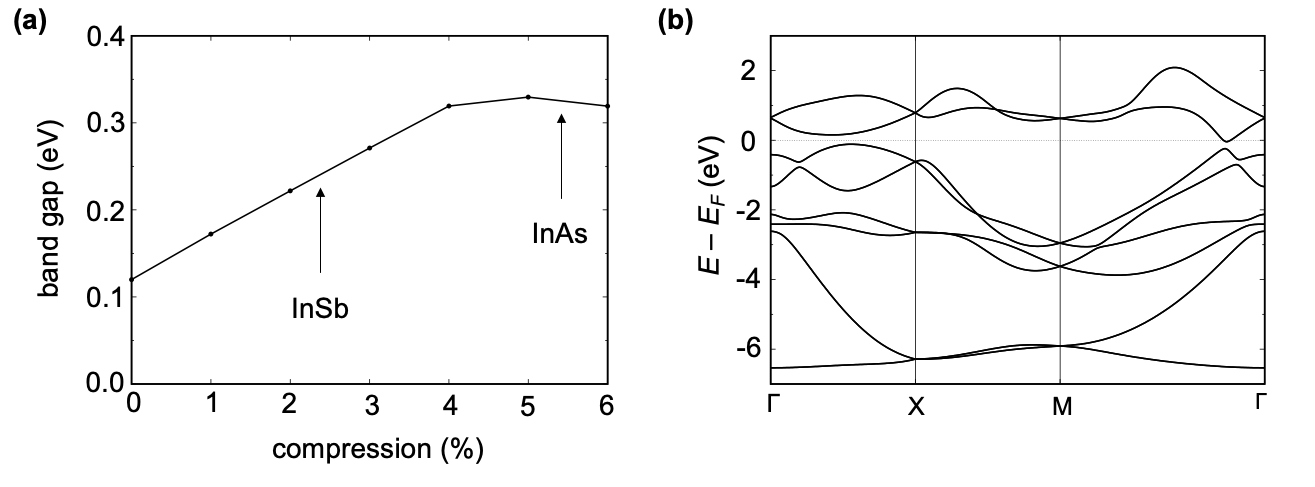}
\caption{(a) The direct band gap of the PE InBi monolayer up to 6\% biaxial compression. The lattice constants of InSb and InAs correspond to 2.3\% and 6.0\% compression. (b) The calculated band structures of InBi monolayer under 6\% compression.}.
\label{fig:strain}
\end{figure}

To investigate the effect of biaxial compression (which preserves inversion symmetry), we calculate the direct band gaps of PE InBi monolayer up to 6\% biaxial compression, shown in Fig.~\ref{fig:strain}(a). As the unit cell is compressed, the band gap increases, which indicates no topological phase transitions during the compression. We further calculate the band structures of the PE InBi monolayer under 6\% compression, shown in Fig.~\ref{fig:strain}(b), and find out that its $\mathrm{Z}_2$ index is 0, consistent with our theory that compression does not lead to topological insulators.

Similar to the orbital analysis presented in the main text, we find out that the crystal-field splitting magnitude, characterized by the energy difference between $p_z$ and $p_{x,y}$ orbitals of the anion atoms, decreases from InAs to InBi. This results from the larger anions and larger distance between the In atomic planes and the anion atomic planes. Since the crystal-field splitting contributes negatively to the band inversion, the band inversion effect is the largest in InBi. 

\begin{figure}[htb]
\centering
\includegraphics[width=\linewidth]{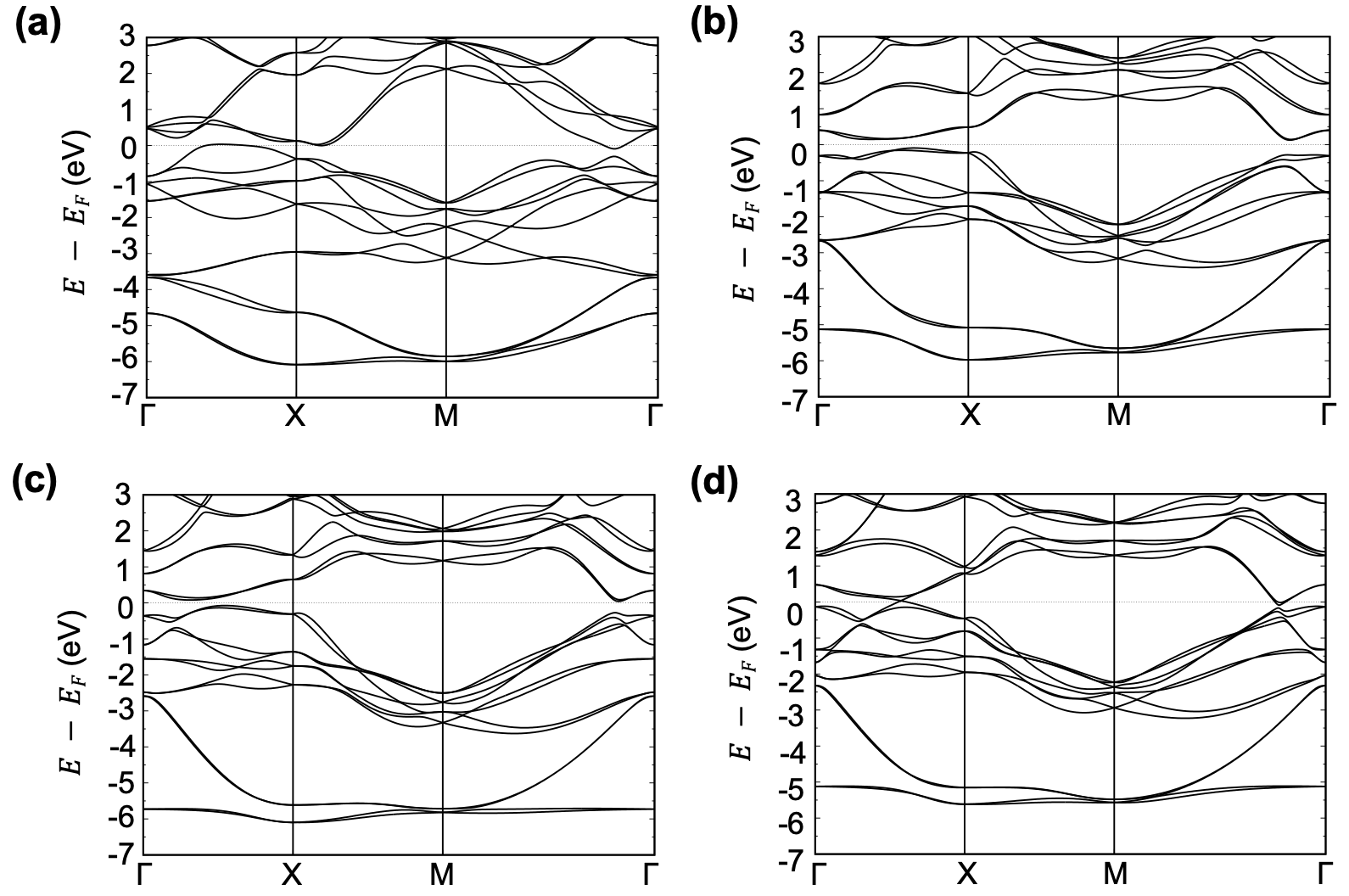}
\caption{The calculated band structures of PE (a) InN$_{0.5}$Bi$_{0.5}$, (b) InP$_{0.5}$Bi$_{0.5}$, (c) InAs$_{0.5}$Bi$_{0.5}$, and (d) InSb$_{0.5}$Bi$_{0.5}$ with spin-orbit coupling.}.
\label{fig:all_alloy}
\end{figure}

Furthermore, we calculated the electronic structures of the alloys InN$_{0.5}$Bi$_{0.5}$, InP$_{0.5}$Bi$_{0.5}$, InAs$_{0.5}$Bi$_{0.5}$, and InSb$_{0.5}$Bi$_{0.5}$ with spin-orbit coupling, shown in Fig.~\ref{fig:all_alloy}. Among them, only InSb$_{0.5}$Bi$_{0.5}$ is not a topological insulator, while the other alloys are topological insulators. This is because the onsite potential energy difference between Sb and Bi is not strong enough to induce the topological phase transition.

\section{Electronic Structures of FE-II and FE-III Monolayers} \label{app:FE}
\begin{figure}[htb]
\centering
\includegraphics[width=\linewidth]{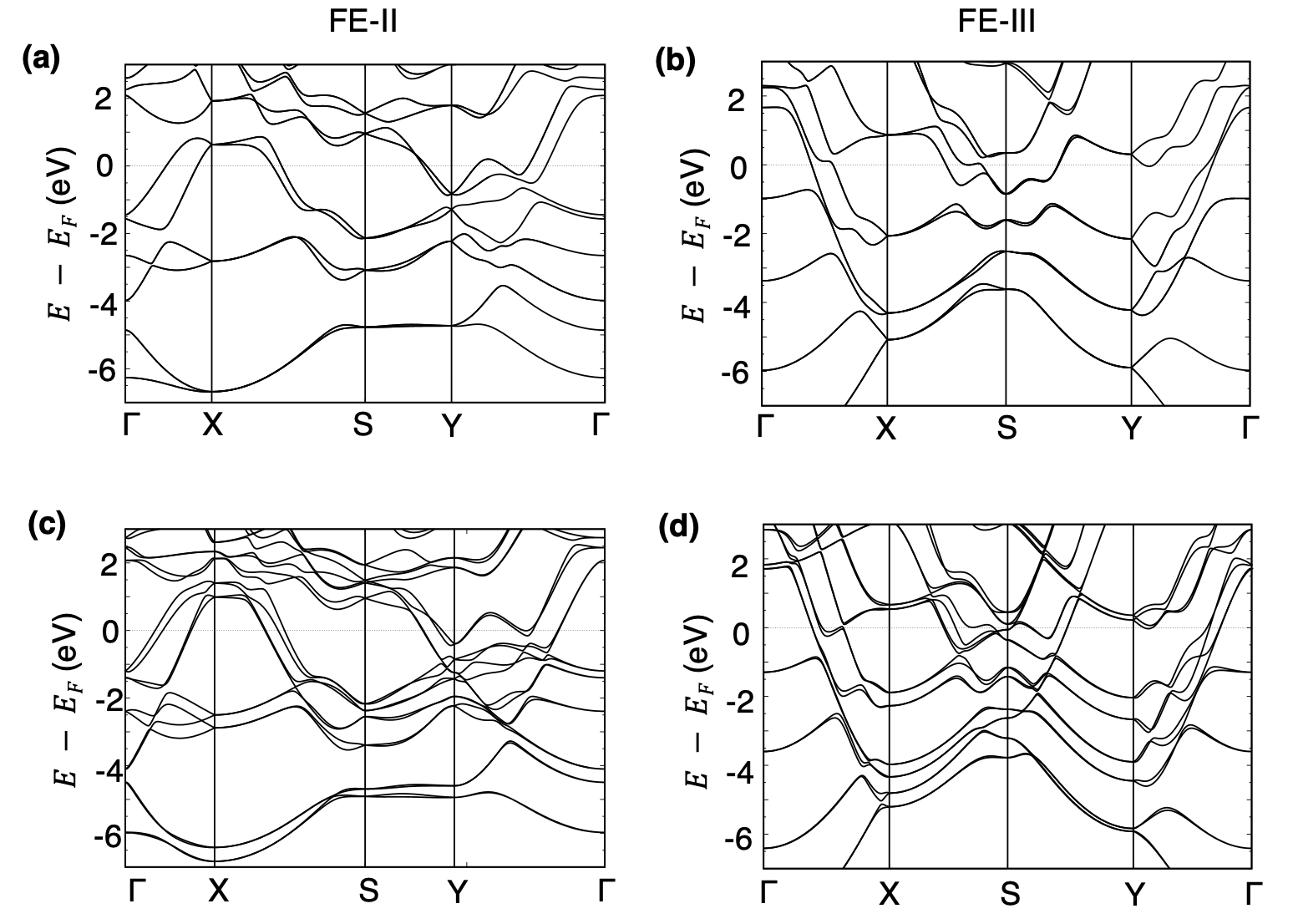}
\caption{(a, b) The calculated band structures of InBi in the (a) FE-II and (b) FE-III phase. (c, d) Same as (a, b), but of InAs$_{0.5}$Bi$_{0.5}$.}
\label{fig:FE-II_III}
\end{figure}

The calculated band structures with spin-orbit coupling of InBi and InAs$_{0.5}$Bi$_{0.5}$ are shown in Fig.~\ref{fig:FE-II_III}. Because the FE-II and FE-III phases are highly compressed along one in-plane axis and adjacent In atoms are strongly buckled, these materials are all metallic and not topological insulators.

\bibliographystyle{apsrev4-1}
\bibliography{reference}
\end{document}